\documentclass[aps,pra,twocolumn,notitlepage]{revtex4-1}
\usepackage{amsmath,amssymb,graphicx,mathtools,amsbsy}
\usepackage[T1]{fontenc}
\usepackage{soul}
\usepackage{color}
\begin{document}

\newcommand{\bra}[1]    {\langle #1|}
\newcommand{\ket}[1]    {|#1 \rangle}
\newcommand{\ketbra}[2]{|#1\rangle\!\langle#2|}
\newcommand{\tr}[1]    {{\rm Tr}\left[ #1 \right]}
\newcommand{\av}[1]    {\langle #1 \rangle}
\newcommand{\en}{\mathcal E_N}

\title{Many-body nonlocality as a resource for quantum-enhanced metrology}

\author{Artur Niezgoda and Jan Chwede\'nczuk}
\affiliation{Faculty of Physics, University of Warsaw, ul. Pasteura 5, PL--02--093 Warszawa, Poland}

\begin{abstract}
  We demonstrate that a many-body nonlocality  is a resource for ultra-precise metrology. This result is achieved by linking the sensitivity of a quantum sensor
  with a combination of many-body correlation functions that witness the nonlocality. We illustrate our findings with some prominent examples---a collection of spins
  forming an Ising chain and a gas of ultra-cold atoms in any two-mode configuration.
\end{abstract}
\maketitle

Entanglement~\cite{ent_rmp}, the Einstein-Podolsky-Rosen steering~\cite{epr,steering,steering2} and the Bell nonlocality~\cite{bell,bell_local} 
play a pivotal role in our understanding of quantum mechanics and can enhance various protocols, 
for instance in quantum cryptography~\cite{PhysRevLett.67.661,qkd0,qkd1,qkd2,qkd3} or quantum computing~\cite{comp1,comp2}.
Entanglement is the main resource for sub shot-noise metrology~\cite{giovannetti2004quantum,pezze2009entanglement} and recently it has been demonstrated that it
is possible to EPR-steer a quantum
sensor to improve its sensitivity~\cite{yadin2020quantum}. Finally, the role of
Bell correlations in some specific cases has been analyzed in the context of quantum metrology~\cite{treutlein_bellqfi,PhysRevA.99.062115}.
Bell correlations and nonlocality have been observed with photons~\cite{test1,test2,test3,test4,test5,test6,test7,test8,test9,test10,loophole}, Josephson qubits~\cite{test11} 
or massive particles~\cite{PhysRevD.14.2543,PhysRevLett.119.010402,schmied2016bell,shin2019bell}. On the other hand, quantum-enhanced sensors operating on many-body systems
have been realized in  various configurations~\cite{gross2010nonlinear,riedel2010atom,leroux2010orientation,chen2011conditional,esteve2008squeezing,smerzi_ob}.

Here we demonstrate that a many-body nonlocality is {\it sufficient} to reach
very high sensitivities, i.e., it is a resource for ultra-precise metrology. We derive a lower bound for the quantum Fisher information (QFI), the central object in 
quantum metrology~\cite{braunstein1994statistical},
in terms of a series of many-body Bell correlators and thus link the nonlocality with the performance of a sensor. These results are corroborated by the QFI and the Bell correlators calculated
for exemplary systems of
qubits forming an Ising chain~\cite{RevModPhys.39.883,baxter2016exactly} or the Bose-Einstein condensate in the double well potential~\cite{dalfovo1999theory,shin2004atom,gati2006noise}.
Showing the mutual relation between the nonlocality and metrology may prove important for these two areas of research and for our understanding of the many-body quantum mechanics.

We rely on a model of local realism that takes $N$ parties, 
and each party independently measures two quantities $\sigma^{(k)}_{1/2}$ which give binary ($\pm1$) outcomes. We construct a $N$-party (-body or -particle)
correlator from an average of a product of outcomes of many experiments
\begin{align}\label{eq.def.e}
  \mathcal E_{\vec n_+,\vec n_-}=|\av{\prod_{k=1}^N\sigma^{(k)}_\pm}|^2,
\end{align}
where $\sigma^{(k)}_\pm=\frac12(\sigma_1^{(k)}\pm i\sigma_2^{(k)})$.The labels $\vec n_+$ and $\vec n_-$ inform which $n_+$ parties picked the $+$ or the $-$ sign ($n_-=N-n_+$). 
If this average is consistent with the postulates of local realism, it can be expressed as
\begin{align}\label{eq.bell}
  \mathcal E_{\vec n_+,\vec n_-}&=\vert\int\!d\lambda\,p(\lambda)\prod_{k=1}^N\sigma^{(k)}_\pm(\lambda)\vert^2\\
  &\leqslant\int\!d\lambda\,p(\lambda)\prod_{k=1}^N|\sigma^{(k)}_\pm(\lambda)|^2=2^{-N}\nonumber,
\end{align}
where $\lambda$ is a local hidden variable and $p(\lambda)$ is its probability distribution. In the second line of Eq.~\eqref{eq.bell} we used the Cauchy-Schwarz inequality to derive 
the upper bound for the correlator, $\mathcal E_{\vec n_+,\vec n_-}\leqslant2^{-N}$, which is the many-body Bell inequality well-suited to test the nonlocality 
in multi-qubit systems~\cite{cavalcanti2007bell,he2011entanglement,cavalcanti2011unified}.

Information about the multiparticle entanglement and nonlocality is encoded in a single
element of the density matrix: such that governs the $\uparrow$-$\downarrow$ coherence of the set of $\vec n_+$ qubits and the $\downarrow$-$\uparrow$ coherence of other $\vec n_-$. Since
any element $\varrho_{nm}$ of the density operator is bounded by $|\varrho_{nm}|^2\leqslant\frac14$, we conclude that also $\mathcal E_{\vec n_+,\vec n_-}\leqslant\frac14$.
The value of $\mathcal E_{\vec n_+,\vec n_-}$ carries information about the depth of nonlocality or entanglement. When $\mathcal E_{\vec n_+,\vec n_-}\in]\frac1{2^N},\frac1{2^{N-1}}]$, 
the correlator can be reproduced with a model, where three out of $N$ qubits are Bell-correlated.  When $\mathcal E_{\vec n_+,\vec n_-}\in]\frac1{2^{N-1}},\frac1{2^{N-2}}]$, 
the nonlocality extends over four qubits. Finally, when $\mathcal E_{\vec n_+,\vec n_-}\in]\frac18,\frac14]$, all $N$ qubits are Bell-correlated (see Appendix~\ref{app.bond} and ~\cite{spiny.milosz}). 

We now consider the scenario where the qubits undergo a metrological transformation, parametrized by $\theta$ (such as the relative phase between the arms of an interferometer
or between the two levels in atomic clocks). We demonstrate that the sensitivity $\Delta\theta$ with which $\theta$ can be estimated is related to those 
correlators $\mathcal E_{\vec n_+,\vec n_-}$ with all combinations of $\vec n_+$ and $\vec n_-$. This way, we establish a link between quantum metrology and the nonlocality.

Consider a quantum system (here in its spectral form)
\begin{align}\label{eq.spect}
  \hat\varrho=\sum_jp_j\ketbra{\psi_j}{\psi_j}.
\end{align}
The evolution --- for instance a passage through an interferometer that introduced the dependence on $\theta$ --- reads in the parameter space
\begin{align}
  i\partial_\theta\hat\varrho=[\hat h,\hat\varrho].
\end{align}
All protocols of estimating $\theta$ have the sensitivity $\Delta\theta$ bounded by
\begin{align}
  \Delta\theta\geqslant\frac1{\sqrt{F_q}}.
\end{align} 
This is the Cramer-Rao lower bound and the $F_q$ is the quantum Fisher information (QFI)~\cite{braunstein1994statistical}, which 
expressed in terms of the eigen states and the corresponding eigenvalues of $\hat\varrho$ [see Eq.~\eqref{eq.spect}] reads
\begin{align}
  F_q&=2\sum_{i,j}\frac{(p_i-p_j)^2}{p_i+p_j}|\bra{\psi_i}\hat h\ket{\psi_j}|^2\nonumber\\
  &=2\sum_{i,j}\frac1{p_i+p_j}|\bra{\psi_i}[\hat\varrho,\hat h]\ket{\psi_j}|^2.\label{eq.qfi}
\end{align}
Since $p_i+p_j\leqslant1$, by neglecting the term in the denominator we obtain the lower bound
\begin{align}\label{eq.qfi.int}
  F_q\geqslant2\sum_{i,j}|\bra{\psi_i}[\hat\varrho,\hat h]\ket{\psi_j}|^2,
\end{align}
which leads to (see Appendix~\ref{app.der}). 
\begin{align}\label{eq.lower}
  F_q\geqslant4\left(\tr{\hat\varrho^2\hat h^2}-\tr{(\hat\varrho\hat h)^2}\right).
\end{align}
This expression is a first step towards establishing a link between the sensitivity and the many-body nonlocality, as we argue below.

For many quantum sensors the generator $\hat h$ takes the form of
\begin{align}\label{eq.gen}
  \hat h=\frac12\sum_{k=1}^N\hat\sigma_\xi^{(k)},
\end{align}
where $\hat\sigma_\xi^{(k)}$ is a Pauli matrix of the $k$-th qubit oriented along the axis $\vec\xi=(\xi_x,\xi_y,\xi_z)$, namely
\begin{align}\label{eq.sigma}
  \hat\sigma_\xi^{(k)}=\xi_x\hat\sigma^{(k)}_x+\xi_y\hat\sigma^{(k)}_y+\xi_z\hat\sigma^{(k)}_z,\ \ \  (\vec\xi\,)^2=1.
\end{align}
This collective generator~\eqref{eq.gen} represents a wide family of interferometric transformations.
For instance $\xi=y$ corresponds to the Mach-Zehnder interferometer with light or atoms or a Ramsey interferometric sequence employed in atomic clocks while $\xi=z$ stands for a phase-shift. 

The eigenstates of a single-qubit operator from Eq.~\eqref{eq.sigma} are $\hat\sigma_\xi^{(k)}\ket{\uparrow/\downarrow}_k=\pm1\ket{\uparrow/\downarrow}_k$
and we represent the density matrix with a basis of $N$-qubit states $\ket n$ being a product of such eigenstates,
\begin{align}
  \hat\varrho=\sum_{n,m=0}^{2^N}\varrho_{nm}\ketbra nm.
\end{align}
Here $n,m$ run through all the combinations of $\uparrow$ and $\downarrow$ independently for each qubit. 
The basis state $\ket n$ is an eigenstate of $\hat h$
\begin{align}\label{eq.eig}
  \hat h\ket n=\left(n_\uparrow-\frac N2\right)\ket n,
\end{align}
where $n_\uparrow$ is the number of $\ket\uparrow$ qubits in $\ket n$. Each eigenstate is $\binom N{n_\uparrow}$ times degenerate, which is a consequence of the collective character of the generator
from Eq.~\eqref{eq.gen}. 
Using the property~\eqref{eq.eig}, we obtain the expression for the lower bound of
the QFI from Eq.~\eqref{eq.lower} in the following form (see Appendix~\ref{app.exp})
\begin{align}\label{eq.bound2}
  F_q\geqslant2\sum_{n,m}(n_\uparrow-m_\uparrow)^2|\varrho_{nm}|^2.
\end{align}
We now focus on a single term of this sum and notice that
every $\ket m$ can be obtained from any $\ket n$ by acting a proper number of times with a rising ($n_+$) and lowering ($n_-$) operator, namely
\begin{align}\label{eq.rise}
  \ket{m}=\hat{\mathcal R}_{\vec n_+}\hat{\mathcal L}_{\vec n_-}\ket n
\end{align}
so that $m_\uparrow=n_\uparrow+n_+-n_-$, giving
\begin{align}
  \varrho_{nm}=\bra n\hat\varrho\hat{\mathcal R}_{\vec n_+}\hat{\mathcal L}_{\vec n_-}\ket n.
\end{align}
By $\hat{\mathcal R}$ and $\hat{\mathcal L}$ we denote a product of rising / lowering  operators
\begin{subequations}\label{eq.rl}
  \begin{align}
    \hat{\mathcal R}_{\vec n_+}&=\hat\sigma_+^{(i_1)}\ldots\hat\sigma_+^{(i_{n_+})}\\
    \hat{\mathcal L}_{\vec n_-}&=\hat\sigma_-^{(j_1)}\ldots\hat\sigma_-^{(j_{n_-})},
  \end{align}
\end{subequations}
where for two directions orthogonal to $\vec\xi$,  $\vec\xi_1$ and $\vec\xi_2$, we have
\begin{align}
  \hat\sigma^{(k)}_\pm=\frac12(\hat\sigma^{(k)}_{\xi_1}\pm i\hat\sigma^{(k)}_{\xi_2}).
\end{align}
The vector symbol $\vec n_{\pm}$ in Eq.~\eqref{eq.rise} indicates that $n_\pm$ qubits are risen / lowered and that they form a particular ordered sub-set of all possible choices from $N$ qubits.

\begin{figure}[h!]
 \center
  \includegraphics[width=1\columnwidth]{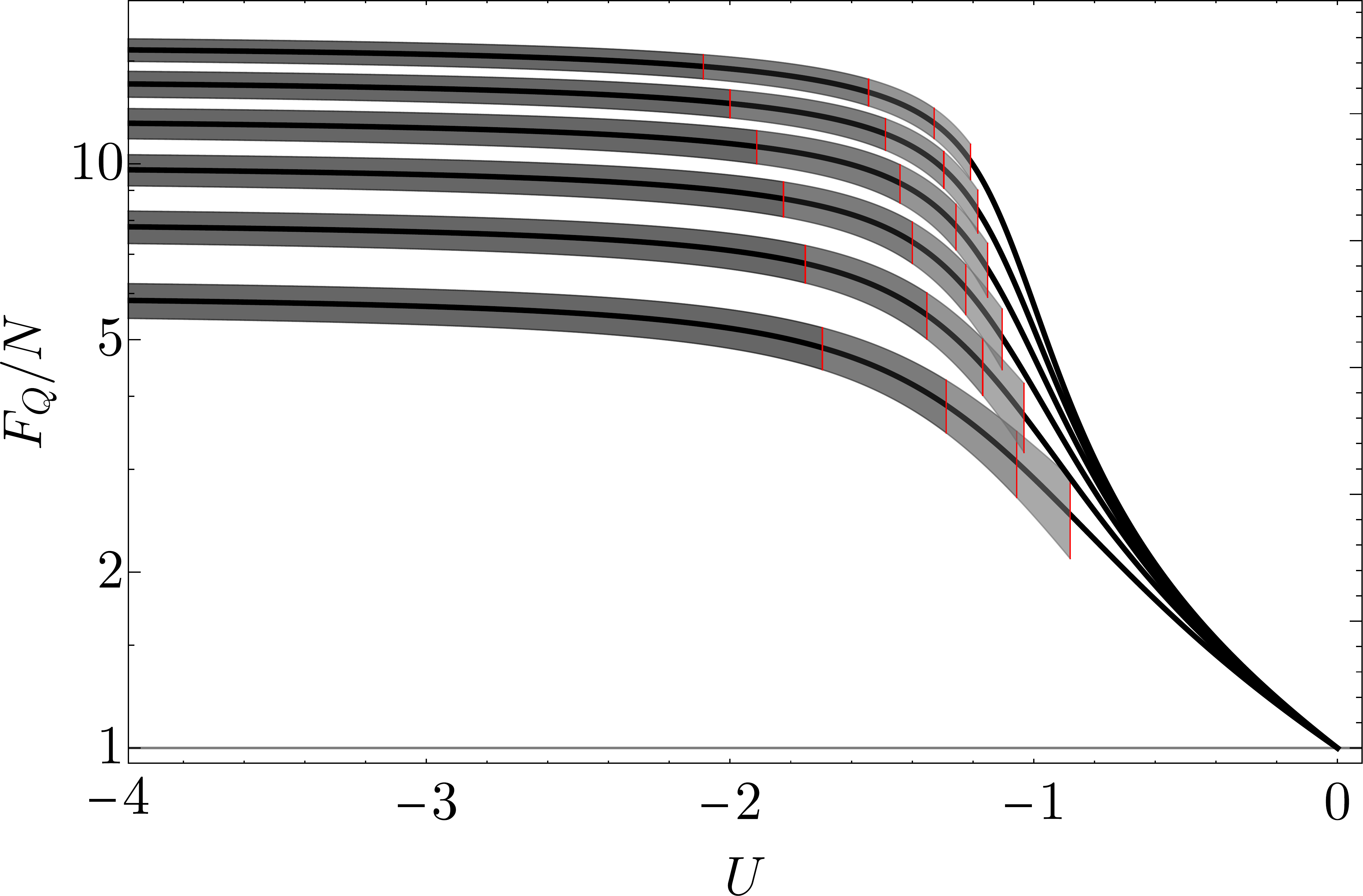}
\centering  
  \caption
      {The QFI calculated with the ground state of the Hamiltonian~\eqref{eq.ising} as a function of $U$ for $N=6,8,10,12,14,16$ (the higher the value reached at the plateau of $F_q=N^2$---i.e., 
        the Heisenberg limit---the bigger the $N$)
        and normalized to the shot noise limit (shown with a grey line). On top of each curve we marked the regions where the correlator $\mathcal E_{N,0}$ detects at least $k$-partite nonlocality.
        Therefore, the darkest patch corresponds to $\mathcal E_{N,0}>\frac18$, when all qubits are Bell-correlated. Next, when $\mathcal E_{N,0}\in]\frac1{16},\frac18]$, so
        the nonlocality encompasses at least $N-1$ qubits, and so forth. The plot shows that the $F_q$ increases monotonously as the number of nonlocally correlated qubits grows. 
        The red lines separating
        the regimes with different strength of nonlocality are added for clarity. }
      \label{fig.chain}
\end{figure}

We now pick all the basis states $\ket n$ and $\ket m$, which are transformed one into another with given $\hat{\mathcal R}_{\vec n_+}$ and $\hat{\mathcal L}_{\vec n_-}$, 
leaving the other $N-(n_++n_-)$ qubits unaltered. We denote such set as $\mathcal A_{\vec n_+,\vec n_-}$. A sum over all such $2^{N-(n_+-n_-)}$
states has a common prefactor $(n_\uparrow-m_\uparrow)^2=(n_+-n_-)^2$. Note that if not for the modulus square in Eq.~\eqref{eq.bound2}, such sum would represent a mean of the product of the two
operators, namely
\begin{align}
  \sum_{n,m\in\mathcal A_{\vec n_+,\vec n_-}}\!\!\!\!\!\!\!\!\!\varrho_{nm}=\tr{\hat\varrho\hat{\mathcal R}_{\vec n_+}\hat{\mathcal L}_{\vec n_-}}.
\end{align}
Using 
\begin{align}\label{eq.ineq4}
  \sum_{i=1}^n|a_i|^2\geqslant\frac1{2^n}|\sum_{i=1}^na_i|^2
\end{align}
which holds for any set of $2^n$ complex numbers (see Appendix~\ref{app.ineq}), we obtain
\begin{align}\label{eq.ineq3}
  \sum_{n,m\in\mathcal A_{\vec n_+,\vec n_-}}\!\!\!\!\!\!\!\!\!|\varrho_{nm}|^2\geqslant\frac1{2^{N-(n_++n_-)}}\mathcal E_{\vec n_+,\vec n_-}
\end{align}
where in correspondence to Eq.~\eqref{eq.def.e} we introduced
\begin{align}
  \mathcal E_{\vec n_+,\vec n_-}\vcentcolon=\big|\tr{\hat\varrho\hat{\mathcal R}_{\vec n_+}\hat{\mathcal L}_{\vec n_-}}\big|^2.
\end{align}
We plug the inequality~\eqref{eq.ineq3} into~\eqref{eq.bound2} and first sum over all possible combinations of fixed $n_+$ and $n_-$, and finally over all $n_+$ and $n_-$, obtaining
the central expression of this work
\begin{align}\label{eq.final}
  F_q\geqslant2\sum_{n_+=0}^N\sum_{n_-=0}^{N-n_+}\frac{(n_+-n_-)^2}{2^{N-(n_++n_-)}}\sum_{\vec n_+,\vec n_-}\mathcal E_{\vec n_+,\vec n_-}.
\end{align}
Thus the QFI and hence the metrological sensitivity is lower-bounded by a combination of $\mathcal E_{\vec n_+,\vec n_-}$, i.e, non-negative Bell correlators of all orders, with non-negative coefficients. 
For a pure separable state 
\begin{align}
  \ket\psi=\bigotimes_{k=1}^N\frac1{\sqrt2}\left(\ket\uparrow_k+\ket\downarrow_k\right),
\end{align}
we have $\mathcal E_{\vec n_+,\vec n_-}=\left(\frac14\right)^{n_++n_-}$ for all $\vec n_+$ and $\vec n_-$ (this is a consequence of the spin-permutation symmetry of this state). 
Hence, the inequality~\eqref{eq.ineq4} 
[and thus~\eqref{eq.ineq3}] is saturated. With this  $\mathcal E_{\vec n_+,\vec n_-}$ the sum over $\vec n_\pm$ in  Eq.~\eqref{eq.final} can be evaluated, 
giving the shot-noise scaling of the QFI with the number of qubits 
$F_q=N$ [note that for pure states, also the inequality~\eqref{eq.qfi.int} is saturated, hence the ``='' sign]. 
To beat the SNL, it is sufficient that correlators such as $\mathcal E_{\vec n_+,\vec n_-}$ grow by any amount from the entanglement--threshold value
$\mathcal E_{\vec n_+,\vec n_-}=\left(\frac14\right)^{n_++n_-}$~\cite{he2011entanglement,spiny.milosz}, 
though not necessarily crossing the Bell limit $\mathcal E_{\vec n_+,\vec n_-}=\left(\frac12\right)^{n_++n_-}$.

However, many-body nonlocality is sufficient (therefore it is a resource) to give ultra-high sensitivity. 
If $\mathcal E_{N,0}>\frac14\frac1{2^{m+1}}$, at least $N-m$ qubits are Bell correlated (see Appendix~\ref{app.bond}), and then Eq.~\eqref{eq.final} gives
\begin{align}\label{eq.heis}
  F_q\geqslant\frac{N^2}{2^{m+1}}.
\end{align}
In particular, when all qubits are nonlocally correlated, then
\begin{align}
  F_q\geqslant\frac{N^2}{2}.
\end{align}
The extreme example is the Greenberger-Horne-Zeilinger (GHZ) state
\begin{align}\label{eq.ghz}
  \ket\psi=\frac1{\sqrt2}\left(\bigotimes_{k=1}^N\ket\uparrow_k+\bigotimes_{k=1}^N\ket\downarrow_k\right),
\end{align}
which gives $\mathcal E_{N,0}=\mathcal E_{0,N}=\frac14$, (all $N$ qubits are Bell-correlated) and $F_q=N^2$.

\begin{figure}[t!]
 \center
  \includegraphics[width=1\columnwidth]{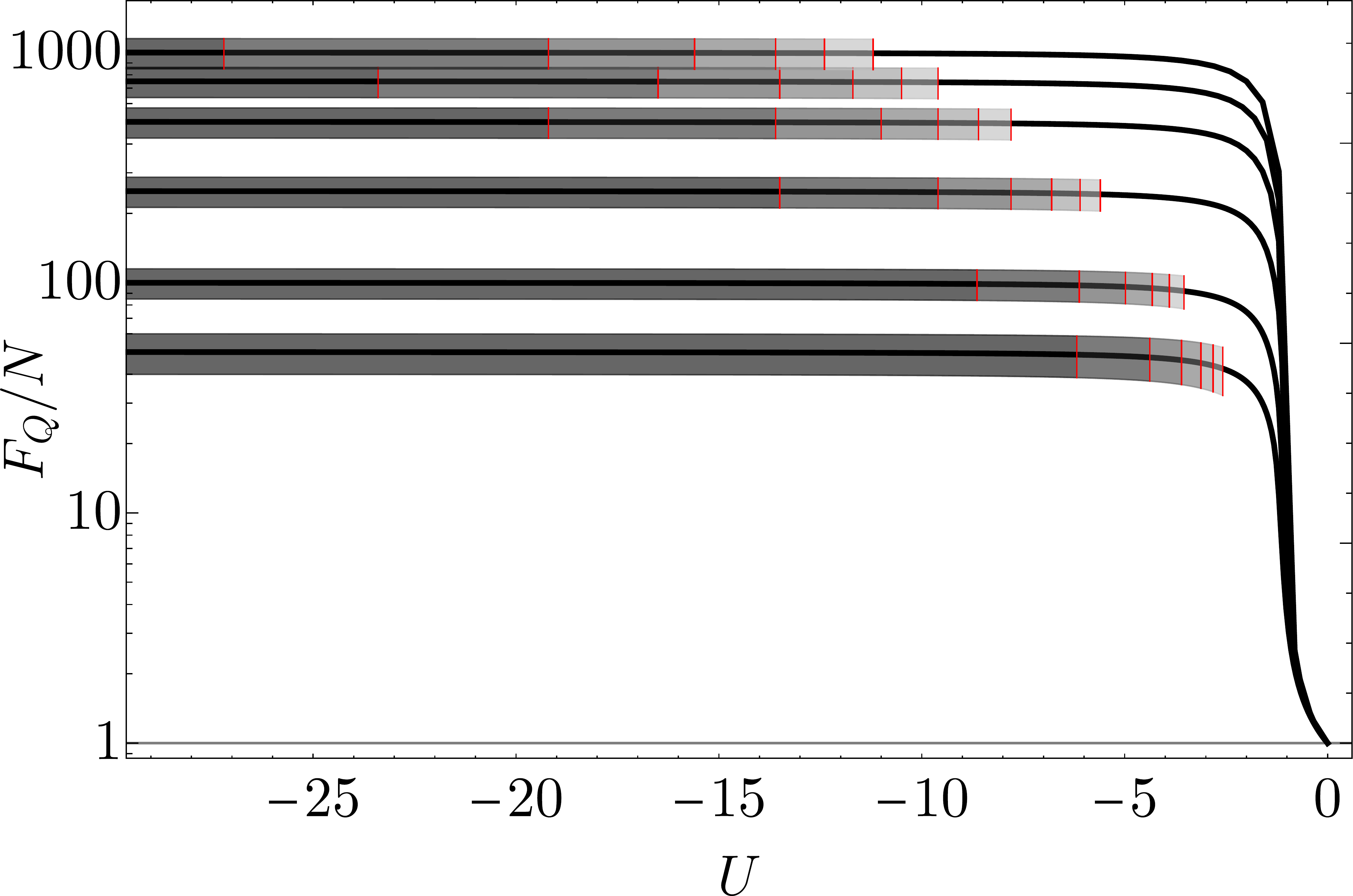}
\centering  
  \caption
      {The QFI calculated with the ground state of the Hamiltonian~\eqref{eq.dw} as a function of $U$ for $N=50, 100, 250, 500, 750, 1000$ (the higher the plateau the bigger the $N$)
        and normalized to the shot noise limit. On top of each curve, the full $N$-body correlator of highest orders is presented with shades of gray, analogically to Fig.~\ref{fig.chain}.
      For higher $N$, the Heisenberg level is approached even when $n<N$ qubits is nonlocally correlated. }
      \label{fig.dw}
\end{figure}

We illustrate these general considerations with some physical examples. First, we take the anti-ferromagnetic Ising Hamiltonian with open boundary conditions, i.e., 
\begin{align}\label{eq.ising}
  \hat H=U\sum_{j=1}^{N-1}\hat\sigma_z^{(j)}\hat\sigma_z^{(j+1)}-\sum_{j=1}^N\hat\sigma_x^{(j)}
\end{align}
where $U$ is the strength of the two-body interactions. We take $N=6,8,10,12,14$ and 16 and for each $N$ find the ground state for different $U<0$. For each $U$ we calculate
the QFI using the formula~\eqref{eq.qfi} with the generator~\eqref{eq.gen}  
aligned along the $\xi=z$ axis. On top of this, we evaluate the $N$-body correlator $\mathcal E_{N,0}$ and highlight the values of $U$ for which 
the correlator detects the many-body nonlocality of the growing order, see Fig.~\ref{fig.chain}. Clearly, the growing depth of nonlocality is linked with approaching the Heisenberg limit, in accordance to Eq.~\eqref{eq.heis}.

As another prominent example we take the collection of $N$ interacting bosonic qubits, such as an ultra-cold Bose gas in a double-well trap. In the two-mode approximation, such a system
can be depicted with the Hamiltonian
\begin{align}\label{eq.dw}
  \hat H=-\hat J_x+U\hat J_z^2,
\end{align}
where the collective spin operators $\hat J_x$ and $\hat J_z$ are given by Eq.~\eqref{eq.gen} with $\xi=x$ and $\xi=z$ respectively. The ground state of this system
undergoes a quantum phase transition as $U$ passes and drops below $-1$~\cite{dziarmaga2002dynamics,trenkwalder2016quantum}, 
and when $U\rightarrow-\infty$ the ground state approaches the GHZ state~\eqref{eq.ghz}, which is well-suited for our purposes.
Figure~\ref{fig.dw} shows the QFI as a function of $U$ for $N=50, 100, 250, 500, 750, 1000$ and the correlator $\mathcal E_{N,0}$. Again, we observe that
when $F_q\simeq N^2$, the system is highly nonlocal. 
When $N\gg1$, the $F_q\simeq N^2$ plateau is reached even when $n<N$ qubits are nonlocally correlated because for large $N$, when a small number of qubits
remains uncorrelated, the coefficient $(n_+-n_-)^2$ from Eq.~\eqref{eq.final} is still close to $N^2$. This shows that for $N\rightarrow\infty$, while the many-body nonlocality remains
sufficient to have high sensitivity, the correlation does not need to encompass all the qubits to have Heisenberg-like scaling. 

\begin{figure}[t!]
 \center
  \includegraphics[width=1\columnwidth]{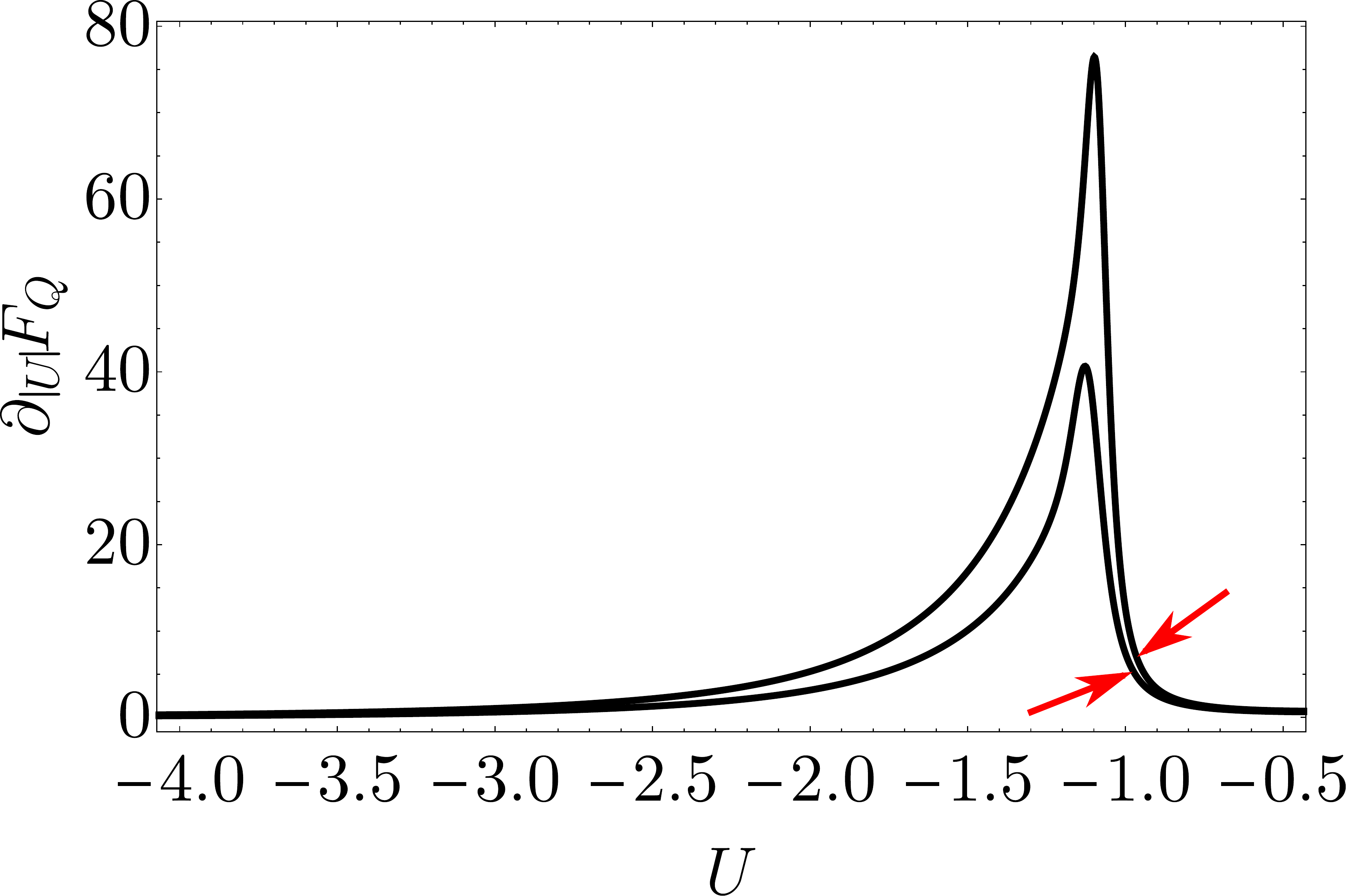}
\centering  
  \caption
      {The derivative of QFI with respect to $U$ for $N=100$ (bottom) and $150$ (top), 
        which shows a rapid growth of sensitivity around the point of quantum phase transition. The red arrows indicate the value of $U$ at which
        the Bell correlator starts to witness the nonlocality, $\mathcal E_{N,0}>\frac1{2^N}$.}
      \label{fig.der}
\end{figure}

Figure~\ref{fig.der}
shows the behaviour of the derivative of QFI over $|U|$ for $N=100$ and $N=150$. Around $U=-1$ point, the QFI starts to rapidly grow (large derivative) and this is accompanied by the
emergence of Bell correlations, $\mathcal E_{N,0}>\frac1{2^{N}}$, as marked by red arrows. This further underlines the significance of nonlocality for quantum sensing and the peculiarity of the transition point.

In this work we have shown that a many-body nonlocality is a driving mechanism for ultra-precise metrology. 
We have expressed the quantum Fisher information in terms of a combination of a particular set of correlation functions of all orders---such that witness the nonlocality extending
over many particles. This central result allowed to provide the lower bound for the sensitivity and identify the necessary condition
to reach the Heisenberg scaling of the $F_q$. These general considerations were illustrated with some prominent examples of multi-qubit systems: a collection of spins
forming an Ising chain and a gas of ultra-cold atoms in any two-mode configuration, for instance trapped in a double-well potential. Our findings shed some light on the
highly non-classical properties of many-body systems and their applications.

We acknowledge the support of the  National Science Centre, Poland under the QuantERA programme, Project no. 2017/25/Z/ST2/03039.

\clearpage
\onecolumngrid
\appendix

\section{Bounds for the correlator}\label{app.bond}

The maximal value of the $N$-party Bell correlator is  $\mathcal E_N=\frac14$~\cite{spiny.milosz} (here we do not specify the number of risen/lowered spins, for clarity of notation). If a single party (say, the first) is not nonlocally with the other, then
\begin{align}\label{eq.app.corr}
  \mathcal E_N=|\av{\prod_{k=1}^N\sigma^{(k)}_\pm}|^2\leqslant\int\! d\lambda\,p(\lambda)\,\mathcal E_{N-1}(\lambda)|\sigma^{(1)}_\pm(\lambda)|^2,
\end{align}
where $\mathcal E_{N-1}$ is the $N-1$-party correlator, while $|\sigma^{(1)}_\pm(\lambda)|^2$ is calculated within the first subsystem. Since also $\mathcal E_{N-1}(\lambda)\leqslant\frac14$,
then the correlator~\eqref{eq.app.corr} is upper-bounded by $\frac18$. This procedure can be continued, and so when 2 parties are not nonlocally correlated, then $\mathcal E_N\leqslant\frac1{16}$.
From this sequence of bounds it follows that when $\mathcal E_N>\frac18$, all parties are Bell-correlated. If $\mathcal E_N>\frac1{16}$, at least $N-1$ are nonlocally correlated, and generally
when $\mathcal E_{N}>\frac14\frac1{2^{m+1}}$, then at least $N-m$ qubits are Bell-correlated.

\section{Derivation of the lower bound~\eqref{eq.lower}}\label{app.der}

Starting from Eq.~\eqref{eq.qfi.int} from the main text, we sum over $j$ and obtain
\begin{align}
  F_q&\geqslant2\sum_i\bra{\psi_i}[\hat\varrho,\hat h][\hat\varrho,\hat h]^\dagger\ket{\psi_i}=2\sum_i\bra{\psi_i}(\hat\varrho\hat h-\hat h\hat\varrho)(\hat h\hat\varrho-\hat\varrho\hat h)\ket{\psi_i}=
  2\sum_i\bra{\psi_i}(\hat\varrho\hat h^2\hat\varrho+\hat h\hat\varrho^2\hat h-\hat\varrho\hat h\hat\varrho\hat h-\hat h\hat\varrho\hat h\hat\varrho)\ket{\psi_i}.
\end{align}
Now we consider each step separately. 
\begin{subequations}
  \begin{align}
    &\sum_i\bra{\psi_i}\hat\varrho\hat h^2\hat\varrho\ket{\psi_i}=\sum_ip_i^2\bra{\psi_i}\hat h^2\ket{\psi_i}=\tr{\hat\varrho^2\hat h^2}.\\
    &\sum_i\bra{\psi_i}\hat h\hat\varrho^2\hat h\ket{\psi_i}=\sum_i\bra{\psi_i}\hat h\sum_jp_j^2\ketbra{\psi_j}{\psi_j}\hat h\ket{\psi_i}=
    \sum_jp_j^2\bra{\psi_j}\hat h^2\ket{\psi_j}=\tr{\hat\varrho^2\hat h^2}.\\
    &\sum_i\bra{\psi_i}\hat\varrho\hat h\hat\varrho\hat h\ket{\psi_i}=\tr{(\hat\varrho\hat h)^2}\\
    &\sum_i\bra{\psi_i}\hat h\hat\varrho\hat h\hat\varrho\ket{\psi_i}=\tr{\hat h\hat\varrho\hat h\hat\varrho}=\tr{\hat\varrho\hat h\hat\varrho\hat h}=\tr{(\hat\varrho\hat h)^2}.
  \end{align}
\end{subequations}
Therefore, the lower bound is 
\begin{align}\label{eq.up}
  F_q\geqslant4\left(\tr{\hat\varrho^2\hat h^2}-\tr{(\hat\varrho\hat h)^2}\right)
\end{align}
as quoted in the main text.

\section{Explicit formula for the lower bound}\label{app.exp}
We now consider both terms of the lower bound~\eqref{eq.up} separately. First, we have
  \begin{align}
    \tr{\hat\varrho^2\hat h^2}&=\tr{\hat h\hat\varrho^2\hat h}=\tr{\sum_{n,k,m}\left(n_\uparrow-\frac N2\right)\left(m_\uparrow-\frac N2\right)\varrho_{nk}\varrho_{km}\ketbra{n}{m}}=
    \sum_{n,m}\left(n_\uparrow-\frac N2\right)^2|\varrho_{nm}|^2.
  \end{align}
The second term reads
  \begin{align}
  \tr{(\hat\varrho\hat h)^2}&=\tr{\hat h\hat\varrho\hat h\hat\varrho}=\tr{\sum_{n,m}\left(n_\uparrow-\frac N2\right)\left(m_\uparrow-\frac N2\right)\varrho_{nm}\ketbra{n}{m}
    \sum_{k,l}\varrho_{kl}\ketbra{k}{l}}\nonumber\\
  &=\sum_{n,m}\left(n_\uparrow-\frac N2\right)\left(m_\uparrow-\frac N2\right)|\varrho_{nm}|^2.
  \end{align}
Therefore, the lower bound for the QFI reads
  \begin{align}
    F_q\geqslant4\sum_{n,m}\left[\left(n_\uparrow-\frac N2\right)^2-\left(n_\uparrow-\frac N2\right)\left(m_\uparrow-\frac N2\right)\right]|\varrho_{nm}|^2=
    4\sum_{n,m}\left(n_\uparrow-\frac N2\right)(n_\uparrow-m_\uparrow)|\varrho_{nm}|^2.
  \end{align}
Now, by splitting this sum into two equal parts and in one of them exchanging the indices $n\leftrightarrow m$, we obtain
  \begin{align}
    F_q\geqslant2\sum_{n,m}\left(n_\uparrow-\frac N2\right)(n_\uparrow-m_\uparrow)|\varrho_{nm}|^2+2\sum_{n,m}\left(m_\uparrow-\frac N2\right)(m_\uparrow-n_\uparrow)|\varrho_{mn}|^2.
  \end{align}
Since $|\varrho_{mn}|^2=|\varrho_{nm}|^2$ we obtain the final expression
\begin{align}
  F_q\geqslant2\sum_{n,m}(n_\uparrow-m_\uparrow)^2|\varrho_{nm}|^2.
\end{align}

\section{Proof for the algebraic inequality}\label{app.ineq}
Here we prove that
\begin{align}\label{eq.ineq1}
  \sum_{i=1}^{2^n}|a_i|^2\geqslant\frac1{2^n}\Big|\sum_{i=1}^{2^n}a_i\Big|^2,
\end{align}
for any set of complex numbers $\{a_1\ldots a_{2^n}\}$. To this end, note that for $n=1$ we have
\begin{align}\label{eq.ineq2}
  |a_1|^2+|a_2|^2=\frac12\left(|a_1+a_2|^2+|a_1-a_2|^2\right)\geqslant\frac12|a_1+a_2|^2.
\end{align}
Similarly, for $n=2$
  \begin{align}
    &|a_1|^2+|a_2|^2+|a_3|^2+|a_4|^2=\nonumber\\
    &=\frac14\left(|a_1+a_2+(a_3+a_4)|^2+|a_1+a_2-(a_3+a_4)|^2+|a_1-a_2+(a_3-a_4)|^2+|a_1-a_2-(a_3-a_4)|^2\right)\nonumber\\
    &\geqslant\frac14|a_1+a_2+a_3+a_4|^2.
  \end{align}
It is now clear how to proceed for higher $n$. First, the full expression is divided into two halves, namely
\begin{align}
  b_1=\sum_{i=1}^{2^{n-1}}a_i,\ \ \ b_2=\sum_{i=1}^{2^{n-1}}a_{i+2^{n-1}}
\end{align}
and we apply the formula~\eqref{eq.ineq2} with $a_1$ and $a_2$ replaced with $b_1$ and $b_2$. Next, a minus sign must be inserted in the middle of $b_1$ and $b_2$, giving
\begin{align}
  &c_1=\sum_{i=1}^{2^{n-2}}a_i-\sum_{i=1}^{2^{n-2}}a_{i+2^{n-2}},\\
  &c_2=\sum_{i=1}^{2^{n-2}}a_{i+2^{n-1}}-\sum_{i=1}^{2^{n-2}}a_{i+2^{n-1}+2^{n-2}}.
\end{align}
This procedure is iterated a total of $2^n$ times, giving $2^n$ terms
\begin{align}
  \sum_{i=1}^{2^n}|a_i|^2=\frac1{2^n}(&|b_1+b_2|^2+|b_1-b_2|^2\\
  &+|c_1+c_2|^2+|c_1-c_2|^2+\ldots).
\end{align}
By neglecting all the terms apart from the first one, we obtain the inequality~\eqref{eq.ineq1}.


\begin{thebibliography}{52}%
\makeatletter
\providecommand \@ifxundefined [1]{%
 \@ifx{#1\undefined}
}%
\providecommand \@ifnum [1]{%
 \ifnum #1\expandafter \@firstoftwo
 \else \expandafter \@secondoftwo
 \fi
}%
\providecommand \@ifx [1]{%
 \ifx #1\expandafter \@firstoftwo
 \else \expandafter \@secondoftwo
 \fi
}%
\providecommand \natexlab [1]{#1}%
\providecommand \enquote  [1]{``#1''}%
\providecommand \bibnamefont  [1]{#1}%
\providecommand \bibfnamefont [1]{#1}%
\providecommand \citenamefont [1]{#1}%
\providecommand \href@noop [0]{\@secondoftwo}%
\providecommand \href [0]{\begingroup \@sanitize@url \@href}%
\providecommand \@href[1]{\@@startlink{#1}\@@href}%
\providecommand \@@href[1]{\endgroup#1\@@endlink}%
\providecommand \@sanitize@url [0]{\catcode `\\12\catcode `\$12\catcode
  `\&12\catcode `\#12\catcode `\^12\catcode `\_12\catcode `\%12\relax}%
\providecommand \@@startlink[1]{}%
\providecommand \@@endlink[0]{}%
\providecommand \url  [0]{\begingroup\@sanitize@url \@url }%
\providecommand \@url [1]{\endgroup\@href {#1}{\urlprefix }}%
\providecommand \urlprefix  [0]{URL }%
\providecommand \Eprint [0]{\href }%
\providecommand \doibase [0]{http://dx.doi.org/}%
\providecommand \selectlanguage [0]{\@gobble}%
\providecommand \bibinfo  [0]{\@secondoftwo}%
\providecommand \bibfield  [0]{\@secondoftwo}%
\providecommand \translation [1]{[#1]}%
\providecommand \BibitemOpen [0]{}%
\providecommand \bibitemStop [0]{}%
\providecommand \bibitemNoStop [0]{.\EOS\space}%
\providecommand \EOS [0]{\spacefactor3000\relax}%
\providecommand \BibitemShut  [1]{\csname bibitem#1\endcsname}%
\let\auto@bib@innerbib\@empty
\bibitem [{\citenamefont {Horodecki}\ \emph {et~al.}(2009)\citenamefont
  {Horodecki}, \citenamefont {Horodecki}, \citenamefont {Horodecki},\ and\
  \citenamefont {Horodecki}}]{ent_rmp}%
  \BibitemOpen
  \bibfield  {author} {\bibinfo {author} {\bibfnamefont {R.}~\bibnamefont
  {Horodecki}}, \bibinfo {author} {\bibfnamefont {P.}~\bibnamefont
  {Horodecki}}, \bibinfo {author} {\bibfnamefont {M.}~\bibnamefont
  {Horodecki}}, \ and\ \bibinfo {author} {\bibfnamefont {K.}~\bibnamefont
  {Horodecki}},\ }\href@noop {} {\bibfield  {journal} {\bibinfo  {journal}
  {Rev. Mod. Phys.}\ }\textbf {\bibinfo {volume} {81}},\ \bibinfo {pages} {865}
  (\bibinfo {year} {2009})}\BibitemShut {NoStop}%
\bibitem [{\citenamefont {Einstein}\ \emph {et~al.}(1935)\citenamefont
  {Einstein}, \citenamefont {Podolsky},\ and\ \citenamefont {Rosen}}]{epr}%
  \BibitemOpen
  \bibfield  {author} {\bibinfo {author} {\bibfnamefont {A.}~\bibnamefont
  {Einstein}}, \bibinfo {author} {\bibfnamefont {B.}~\bibnamefont {Podolsky}},
  \ and\ \bibinfo {author} {\bibfnamefont {N.}~\bibnamefont {Rosen}},\
  }\href@noop {} {\bibfield  {journal} {\bibinfo  {journal} {Phys. Rev.}\
  }\textbf {\bibinfo {volume} {47}},\ \bibinfo {pages} {777} (\bibinfo {year}
  {1935})}\BibitemShut {NoStop}%
\bibitem [{\citenamefont {He}\ \emph {et~al.}(2012)\citenamefont {He},
  \citenamefont {Drummond}, \citenamefont {Olsen},\ and\ \citenamefont
  {Reid}}]{steering}%
  \BibitemOpen
  \bibfield  {author} {\bibinfo {author} {\bibfnamefont {Q.~Y.}\ \bibnamefont
  {He}}, \bibinfo {author} {\bibfnamefont {P.~D.}\ \bibnamefont {Drummond}},
  \bibinfo {author} {\bibfnamefont {M.~K.}\ \bibnamefont {Olsen}}, \ and\
  \bibinfo {author} {\bibfnamefont {M.~D.}\ \bibnamefont {Reid}},\ }\href@noop
  {} {\bibfield  {journal} {\bibinfo  {journal} {Phys. Rev. A}\ }\textbf
  {\bibinfo {volume} {86}},\ \bibinfo {pages} {023626} (\bibinfo {year}
  {2012})}\BibitemShut {NoStop}%
\bibitem [{\citenamefont {Cavalcanti}\ \emph {et~al.}(2009)\citenamefont
  {Cavalcanti}, \citenamefont {Jones}, \citenamefont {Wiseman},\ and\
  \citenamefont {Reid}}]{steering2}%
  \BibitemOpen
  \bibfield  {author} {\bibinfo {author} {\bibfnamefont {E.~G.}\ \bibnamefont
  {Cavalcanti}}, \bibinfo {author} {\bibfnamefont {S.~J.}\ \bibnamefont
  {Jones}}, \bibinfo {author} {\bibfnamefont {H.~M.}\ \bibnamefont {Wiseman}},
  \ and\ \bibinfo {author} {\bibfnamefont {M.~D.}\ \bibnamefont {Reid}},\
  }\href@noop {} {\bibfield  {journal} {\bibinfo  {journal} {Phys. Rev. A}\
  }\textbf {\bibinfo {volume} {80}},\ \bibinfo {pages} {032112} (\bibinfo
  {year} {2009})}\BibitemShut {NoStop}%
\bibitem [{\citenamefont {Bell}(1964)}]{bell}%
  \BibitemOpen
  \bibfield  {author} {\bibinfo {author} {\bibfnamefont {J.~S.}\ \bibnamefont
  {Bell}},\ }\href@noop {} {\bibfield  {journal} {\bibinfo  {journal}
  {Physics}\ }\textbf {\bibinfo {volume} {1}},\ \bibinfo {pages} {195}
  (\bibinfo {year} {1964})}\BibitemShut {NoStop}%
\bibitem [{\citenamefont {Brunner}\ \emph {et~al.}(2014)\citenamefont
  {Brunner}, \citenamefont {Cavalcanti}, \citenamefont {Pironio}, \citenamefont
  {Scarani},\ and\ \citenamefont {Wehner}}]{bell_local}%
  \BibitemOpen
  \bibfield  {author} {\bibinfo {author} {\bibfnamefont {N.}~\bibnamefont
  {Brunner}}, \bibinfo {author} {\bibfnamefont {D.}~\bibnamefont {Cavalcanti}},
  \bibinfo {author} {\bibfnamefont {S.}~\bibnamefont {Pironio}}, \bibinfo
  {author} {\bibfnamefont {V.}~\bibnamefont {Scarani}}, \ and\ \bibinfo
  {author} {\bibfnamefont {S.}~\bibnamefont {Wehner}},\ }\href@noop {}
  {\bibfield  {journal} {\bibinfo  {journal} {Rev. Mod. Phys.}\ }\textbf
  {\bibinfo {volume} {86}},\ \bibinfo {pages} {419} (\bibinfo {year}
  {2014})}\BibitemShut {NoStop}%
\bibitem [{\citenamefont {Ekert}(1991)}]{PhysRevLett.67.661}%
  \BibitemOpen
  \bibfield  {author} {\bibinfo {author} {\bibfnamefont {A.~K.}\ \bibnamefont
  {Ekert}},\ }\href {\doibase 10.1103/PhysRevLett.67.661} {\bibfield  {journal}
  {\bibinfo  {journal} {Phys. Rev. Lett.}\ }\textbf {\bibinfo {volume} {67}},\
  \bibinfo {pages} {661} (\bibinfo {year} {1991})}\BibitemShut {NoStop}%
\bibitem [{\citenamefont {Chun-Yan}\ \emph {et~al.}(2005)\citenamefont
  {Chun-Yan}, \citenamefont {Hong-Yu}, \citenamefont {Yan},\ and\ \citenamefont
  {Fu-Guo}}]{qkd0}%
  \BibitemOpen
  \bibfield  {author} {\bibinfo {author} {\bibfnamefont {L.}~\bibnamefont
  {Chun-Yan}}, \bibinfo {author} {\bibfnamefont {Z.}~\bibnamefont {Hong-Yu}},
  \bibinfo {author} {\bibfnamefont {W.}~\bibnamefont {Yan}}, \ and\ \bibinfo
  {author} {\bibfnamefont {D.}~\bibnamefont {Fu-Guo}},\ }\href@noop {}
  {\bibfield  {journal} {\bibinfo  {journal} {Chinese Physics Letters}\
  }\textbf {\bibinfo {volume} {22}},\ \bibinfo {pages} {1049} (\bibinfo {year}
  {2005})}\BibitemShut {NoStop}%
\bibitem [{\citenamefont {Barrett}\ \emph {et~al.}(2005)\citenamefont
  {Barrett}, \citenamefont {Hardy},\ and\ \citenamefont {Kent}}]{qkd1}%
  \BibitemOpen
  \bibfield  {author} {\bibinfo {author} {\bibfnamefont {J.}~\bibnamefont
  {Barrett}}, \bibinfo {author} {\bibfnamefont {L.}~\bibnamefont {Hardy}}, \
  and\ \bibinfo {author} {\bibfnamefont {A.}~\bibnamefont {Kent}},\ }\href
  {\doibase 10.1103/PhysRevLett.95.010503} {\bibfield  {journal} {\bibinfo
  {journal} {Phys. Rev. Lett.}\ }\textbf {\bibinfo {volume} {95}},\ \bibinfo
  {pages} {010503} (\bibinfo {year} {2005})}\BibitemShut {NoStop}%
\bibitem [{\citenamefont {Ac\'{\i}n}\ \emph {et~al.}(2006)\citenamefont
  {Ac\'{\i}n}, \citenamefont {Gisin},\ and\ \citenamefont {Masanes}}]{qkd2}%
  \BibitemOpen
  \bibfield  {author} {\bibinfo {author} {\bibfnamefont {A.}~\bibnamefont
  {Ac\'{\i}n}}, \bibinfo {author} {\bibfnamefont {N.}~\bibnamefont {Gisin}}, \
  and\ \bibinfo {author} {\bibfnamefont {L.}~\bibnamefont {Masanes}},\ }\href
  {\doibase 10.1103/PhysRevLett.97.120405} {\bibfield  {journal} {\bibinfo
  {journal} {Phys. Rev. Lett.}\ }\textbf {\bibinfo {volume} {97}},\ \bibinfo
  {pages} {120405} (\bibinfo {year} {2006})}\BibitemShut {NoStop}%
\bibitem [{\citenamefont {Gisin}\ \emph {et~al.}(2010)\citenamefont {Gisin},
  \citenamefont {Pironio},\ and\ \citenamefont {Sangouard}}]{qkd3}%
  \BibitemOpen
  \bibfield  {author} {\bibinfo {author} {\bibfnamefont {N.}~\bibnamefont
  {Gisin}}, \bibinfo {author} {\bibfnamefont {S.}~\bibnamefont {Pironio}}, \
  and\ \bibinfo {author} {\bibfnamefont {N.}~\bibnamefont {Sangouard}},\ }\href
  {\doibase 10.1103/PhysRevLett.105.070501} {\bibfield  {journal} {\bibinfo
  {journal} {Phys. Rev. Lett.}\ }\textbf {\bibinfo {volume} {105}},\ \bibinfo
  {pages} {070501} (\bibinfo {year} {2010})}\BibitemShut {NoStop}%
\bibitem [{\citenamefont {S\o{}rensen}\ and\ \citenamefont
  {M\o{}lmer}(2000)}]{comp1}%
  \BibitemOpen
  \bibfield  {author} {\bibinfo {author} {\bibfnamefont {A.}~\bibnamefont
  {S\o{}rensen}}\ and\ \bibinfo {author} {\bibfnamefont {K.}~\bibnamefont
  {M\o{}lmer}},\ }\href {\doibase 10.1103/PhysRevA.62.022311} {\bibfield
  {journal} {\bibinfo  {journal} {Phys. Rev. A}\ }\textbf {\bibinfo {volume}
  {62}},\ \bibinfo {pages} {022311} (\bibinfo {year} {2000})}\BibitemShut
  {NoStop}%
\bibitem [{\citenamefont {Chen}\ \emph {et~al.}(2014)\citenamefont {Chen},
  \citenamefont {Menicucci},\ and\ \citenamefont {Pfister}}]{comp2}%
  \BibitemOpen
  \bibfield  {author} {\bibinfo {author} {\bibfnamefont {M.}~\bibnamefont
  {Chen}}, \bibinfo {author} {\bibfnamefont {N.~C.}\ \bibnamefont {Menicucci}},
  \ and\ \bibinfo {author} {\bibfnamefont {O.}~\bibnamefont {Pfister}},\ }\href
  {\doibase 10.1103/PhysRevLett.112.120505} {\bibfield  {journal} {\bibinfo
  {journal} {Phys. Rev. Lett.}\ }\textbf {\bibinfo {volume} {112}},\ \bibinfo
  {pages} {120505} (\bibinfo {year} {2014})}\BibitemShut {NoStop}%
\bibitem [{\citenamefont {Giovannetti}\ \emph {et~al.}(2004)\citenamefont
  {Giovannetti}, \citenamefont {Lloyd},\ and\ \citenamefont
  {Maccone}}]{giovannetti2004quantum}%
  \BibitemOpen
  \bibfield  {author} {\bibinfo {author} {\bibfnamefont {V.}~\bibnamefont
  {Giovannetti}}, \bibinfo {author} {\bibfnamefont {S.}~\bibnamefont {Lloyd}},
  \ and\ \bibinfo {author} {\bibfnamefont {L.}~\bibnamefont {Maccone}},\
  }\href@noop {} {\bibfield  {journal} {\bibinfo  {journal} {Science}\ }\textbf
  {\bibinfo {volume} {306}},\ \bibinfo {pages} {1330} (\bibinfo {year}
  {2004})}\BibitemShut {NoStop}%
\bibitem [{\citenamefont {Pezz{\'e}}\ and\ \citenamefont
  {Smerzi}(2009)}]{pezze2009entanglement}%
  \BibitemOpen
  \bibfield  {author} {\bibinfo {author} {\bibfnamefont {L.}~\bibnamefont
  {Pezz{\'e}}}\ and\ \bibinfo {author} {\bibfnamefont {A.}~\bibnamefont
  {Smerzi}},\ }\href@noop {} {\bibfield  {journal} {\bibinfo  {journal} {Phys.
  Rev. Lett.}\ }\textbf {\bibinfo {volume} {102}},\ \bibinfo {pages} {100401}
  (\bibinfo {year} {2009})}\BibitemShut {NoStop}%
\bibitem [{\citenamefont {Yadin}\ \emph {et~al.}(2020)\citenamefont {Yadin},
  \citenamefont {Fadel},\ and\ \citenamefont {Gessner}}]{yadin2020quantum}%
  \BibitemOpen
  \bibfield  {author} {\bibinfo {author} {\bibfnamefont {B.}~\bibnamefont
  {Yadin}}, \bibinfo {author} {\bibfnamefont {M.}~\bibnamefont {Fadel}}, \ and\
  \bibinfo {author} {\bibfnamefont {M.}~\bibnamefont {Gessner}},\ }\href@noop
  {} {\bibfield  {journal} {\bibinfo  {journal} {arXiv preprint
  arXiv:2009.08440}\ } (\bibinfo {year} {2020})}\BibitemShut {NoStop}%
\bibitem [{\citenamefont {Fr\"owis}\ \emph {et~al.}(2019)\citenamefont
  {Fr\"owis}, \citenamefont {Fadel}, \citenamefont {Treutlein}, \citenamefont
  {Gisin},\ and\ \citenamefont {Brunner}}]{treutlein_bellqfi}%
  \BibitemOpen
  \bibfield  {author} {\bibinfo {author} {\bibfnamefont {F.}~\bibnamefont
  {Fr\"owis}}, \bibinfo {author} {\bibfnamefont {M.}~\bibnamefont {Fadel}},
  \bibinfo {author} {\bibfnamefont {P.}~\bibnamefont {Treutlein}}, \bibinfo
  {author} {\bibfnamefont {N.}~\bibnamefont {Gisin}}, \ and\ \bibinfo {author}
  {\bibfnamefont {N.}~\bibnamefont {Brunner}},\ }\href {\doibase
  10.1103/PhysRevA.99.040101} {\bibfield  {journal} {\bibinfo  {journal} {Phys.
  Rev. A}\ }\textbf {\bibinfo {volume} {99}},\ \bibinfo {pages} {040101}
  (\bibinfo {year} {2019})}\BibitemShut {NoStop}%
\bibitem [{\citenamefont {Niezgoda}\ \emph {et~al.}(2019)\citenamefont
  {Niezgoda}, \citenamefont {Chwede\ifmmode~\acute{n}\else \'{n}\fi{}czuk},
  \citenamefont {Pezz\'e},\ and\ \citenamefont {Smerzi}}]{PhysRevA.99.062115}%
  \BibitemOpen
  \bibfield  {author} {\bibinfo {author} {\bibfnamefont {A.}~\bibnamefont
  {Niezgoda}}, \bibinfo {author} {\bibfnamefont {J.}~\bibnamefont
  {Chwede\ifmmode~\acute{n}\else \'{n}\fi{}czuk}}, \bibinfo {author}
  {\bibfnamefont {L.}~\bibnamefont {Pezz\'e}}, \ and\ \bibinfo {author}
  {\bibfnamefont {A.}~\bibnamefont {Smerzi}},\ }\href {\doibase
  10.1103/PhysRevA.99.062115} {\bibfield  {journal} {\bibinfo  {journal} {Phys.
  Rev. A}\ }\textbf {\bibinfo {volume} {99}},\ \bibinfo {pages} {062115}
  (\bibinfo {year} {2019})}\BibitemShut {NoStop}%
\bibitem [{\citenamefont {Freedman}\ and\ \citenamefont
  {Clauser}(1972)}]{test1}%
  \BibitemOpen
  \bibfield  {author} {\bibinfo {author} {\bibfnamefont {S.~J.}\ \bibnamefont
  {Freedman}}\ and\ \bibinfo {author} {\bibfnamefont {J.~F.}\ \bibnamefont
  {Clauser}},\ }\href@noop {} {\bibfield  {journal} {\bibinfo  {journal} {Phys.
  Rev. Lett.}\ }\textbf {\bibinfo {volume} {28}},\ \bibinfo {pages} {938}
  (\bibinfo {year} {1972})}\BibitemShut {NoStop}%
\bibitem [{\citenamefont {Aspect}\ \emph {et~al.}(1981)\citenamefont {Aspect},
  \citenamefont {Grangier},\ and\ \citenamefont {Roger}}]{test2}%
  \BibitemOpen
  \bibfield  {author} {\bibinfo {author} {\bibfnamefont {A.}~\bibnamefont
  {Aspect}}, \bibinfo {author} {\bibfnamefont {P.}~\bibnamefont {Grangier}}, \
  and\ \bibinfo {author} {\bibfnamefont {G.}~\bibnamefont {Roger}},\
  }\href@noop {} {\bibfield  {journal} {\bibinfo  {journal} {Phys. Rev. Lett.}\
  }\textbf {\bibinfo {volume} {47}},\ \bibinfo {pages} {460} (\bibinfo {year}
  {1981})}\BibitemShut {NoStop}%
\bibitem [{\citenamefont {Aspect}\ \emph {et~al.}(1982)\citenamefont {Aspect},
  \citenamefont {Dalibard},\ and\ \citenamefont {Roger}}]{test3}%
  \BibitemOpen
  \bibfield  {author} {\bibinfo {author} {\bibfnamefont {A.}~\bibnamefont
  {Aspect}}, \bibinfo {author} {\bibfnamefont {J.}~\bibnamefont {Dalibard}}, \
  and\ \bibinfo {author} {\bibfnamefont {G.}~\bibnamefont {Roger}},\
  }\href@noop {} {\bibfield  {journal} {\bibinfo  {journal} {Phys. Rev. Lett.}\
  }\textbf {\bibinfo {volume} {49}},\ \bibinfo {pages} {1804} (\bibinfo {year}
  {1982})}\BibitemShut {NoStop}%
\bibitem [{\citenamefont {Tittel}\ \emph
  {et~al.}(1998{\natexlab{a}})\citenamefont {Tittel}, \citenamefont {Brendel},
  \citenamefont {Gisin}, \citenamefont {Herzog}, \citenamefont {Zbinden},\ and\
  \citenamefont {Gisin}}]{test4}%
  \BibitemOpen
  \bibfield  {author} {\bibinfo {author} {\bibfnamefont {W.}~\bibnamefont
  {Tittel}}, \bibinfo {author} {\bibfnamefont {J.}~\bibnamefont {Brendel}},
  \bibinfo {author} {\bibfnamefont {B.}~\bibnamefont {Gisin}}, \bibinfo
  {author} {\bibfnamefont {T.}~\bibnamefont {Herzog}}, \bibinfo {author}
  {\bibfnamefont {H.}~\bibnamefont {Zbinden}}, \ and\ \bibinfo {author}
  {\bibfnamefont {N.}~\bibnamefont {Gisin}},\ }\href@noop {} {\bibfield
  {journal} {\bibinfo  {journal} {Phys. Rev. A}\ }\textbf {\bibinfo {volume}
  {57}},\ \bibinfo {pages} {3229} (\bibinfo {year}
  {1998}{\natexlab{a}})}\BibitemShut {NoStop}%
\bibitem [{\citenamefont {Tittel}\ \emph
  {et~al.}(1998{\natexlab{b}})\citenamefont {Tittel}, \citenamefont {Brendel},
  \citenamefont {Zbinden},\ and\ \citenamefont {Gisin}}]{test5}%
  \BibitemOpen
  \bibfield  {author} {\bibinfo {author} {\bibfnamefont {W.}~\bibnamefont
  {Tittel}}, \bibinfo {author} {\bibfnamefont {J.}~\bibnamefont {Brendel}},
  \bibinfo {author} {\bibfnamefont {H.}~\bibnamefont {Zbinden}}, \ and\
  \bibinfo {author} {\bibfnamefont {N.}~\bibnamefont {Gisin}},\ }\href@noop {}
  {\bibfield  {journal} {\bibinfo  {journal} {Phys. Rev. Lett.}\ }\textbf
  {\bibinfo {volume} {81}},\ \bibinfo {pages} {3563} (\bibinfo {year}
  {1998}{\natexlab{b}})}\BibitemShut {NoStop}%
\bibitem [{\citenamefont {Weihs}\ \emph {et~al.}(1998)\citenamefont {Weihs},
  \citenamefont {Jennewein}, \citenamefont {Simon}, \citenamefont
  {Weinfurter},\ and\ \citenamefont {Zeilinger}}]{test6}%
  \BibitemOpen
  \bibfield  {author} {\bibinfo {author} {\bibfnamefont {G.}~\bibnamefont
  {Weihs}}, \bibinfo {author} {\bibfnamefont {T.}~\bibnamefont {Jennewein}},
  \bibinfo {author} {\bibfnamefont {C.}~\bibnamefont {Simon}}, \bibinfo
  {author} {\bibfnamefont {H.}~\bibnamefont {Weinfurter}}, \ and\ \bibinfo
  {author} {\bibfnamefont {A.}~\bibnamefont {Zeilinger}},\ }\href@noop {}
  {\bibfield  {journal} {\bibinfo  {journal} {Phys. Rev. Lett.}\ }\textbf
  {\bibinfo {volume} {81}},\ \bibinfo {pages} {5039} (\bibinfo {year}
  {1998})}\BibitemShut {NoStop}%
\bibitem [{\citenamefont {{Pan}}\ \emph {et~al.}(2000)\citenamefont {{Pan}},
  \citenamefont {{Bouwmeester}}, \citenamefont {{Daniell}}, \citenamefont
  {{Weinfurter}},\ and\ \citenamefont {{Zeilinger}}}]{test7}%
  \BibitemOpen
  \bibfield  {author} {\bibinfo {author} {\bibfnamefont {J.-W.}\ \bibnamefont
  {{Pan}}}, \bibinfo {author} {\bibfnamefont {D.}~\bibnamefont
  {{Bouwmeester}}}, \bibinfo {author} {\bibfnamefont {M.}~\bibnamefont
  {{Daniell}}}, \bibinfo {author} {\bibfnamefont {H.}~\bibnamefont
  {{Weinfurter}}}, \ and\ \bibinfo {author} {\bibfnamefont {A.}~\bibnamefont
  {{Zeilinger}}},\ }\href@noop {} {\bibfield  {journal} {\bibinfo  {journal}
  {Nature}\ }\textbf {\bibinfo {volume} {403}},\ \bibinfo {pages} {515}
  (\bibinfo {year} {2000})}\BibitemShut {NoStop}%
\bibitem [{\citenamefont {{Kielpinski}}\ \emph {et~al.}(2001)\citenamefont
  {{Kielpinski}}, \citenamefont {{Meyer}}, \citenamefont {{Sackett}},
  \citenamefont {{Itano}}, \citenamefont {{Monroe}},\ and\ \citenamefont
  {{Wineland}}}]{test8}%
  \BibitemOpen
  \bibfield  {author} {\bibinfo {author} {\bibfnamefont {D.}~\bibnamefont
  {{Kielpinski}}}, \bibinfo {author} {\bibfnamefont {V.}~\bibnamefont
  {{Meyer}}}, \bibinfo {author} {\bibfnamefont {C.~A.}\ \bibnamefont
  {{Sackett}}}, \bibinfo {author} {\bibfnamefont {W.~M.}\ \bibnamefont
  {{Itano}}}, \bibinfo {author} {\bibfnamefont {C.}~\bibnamefont {{Monroe}}}, \
  and\ \bibinfo {author} {\bibfnamefont {D.~J.}\ \bibnamefont {{Wineland}}},\
  }\href@noop {} {\bibfield  {journal} {\bibinfo  {journal} {Nature}\ }\textbf
  {\bibinfo {volume} {409}},\ \bibinfo {pages} {791} (\bibinfo {year}
  {2001})}\BibitemShut {NoStop}%
\bibitem [{\citenamefont {{Gr{\"o}blacher}}\ \emph {et~al.}(2007)\citenamefont
  {{Gr{\"o}blacher}}, \citenamefont {{Paterek}}, \citenamefont {{Kaltenbaek}},
  \citenamefont {{Brukner}}, \citenamefont {{{\.Z}ukowski}}, \citenamefont
  {{Aspelmeyer}},\ and\ \citenamefont {{Zeilinger}}}]{test9}%
  \BibitemOpen
  \bibfield  {author} {\bibinfo {author} {\bibfnamefont {S.}~\bibnamefont
  {{Gr{\"o}blacher}}}, \bibinfo {author} {\bibfnamefont {T.}~\bibnamefont
  {{Paterek}}}, \bibinfo {author} {\bibfnamefont {R.}~\bibnamefont
  {{Kaltenbaek}}}, \bibinfo {author} {\bibfnamefont {{\v C}.}~\bibnamefont
  {{Brukner}}}, \bibinfo {author} {\bibfnamefont {M.}~\bibnamefont
  {{{\.Z}ukowski}}}, \bibinfo {author} {\bibfnamefont {M.}~\bibnamefont
  {{Aspelmeyer}}}, \ and\ \bibinfo {author} {\bibfnamefont {A.}~\bibnamefont
  {{Zeilinger}}},\ }\href@noop {} {\bibfield  {journal} {\bibinfo  {journal}
  {Nature}\ }\textbf {\bibinfo {volume} {446}},\ \bibinfo {pages} {871}
  (\bibinfo {year} {2007})}\BibitemShut {NoStop}%
\bibitem [{\citenamefont {Salart}\ \emph {et~al.}(2008)\citenamefont {Salart},
  \citenamefont {Baas}, \citenamefont {van Houwelingen}, \citenamefont
  {Gisin},\ and\ \citenamefont {Zbinden}}]{test10}%
  \BibitemOpen
  \bibfield  {author} {\bibinfo {author} {\bibfnamefont {D.}~\bibnamefont
  {Salart}}, \bibinfo {author} {\bibfnamefont {A.}~\bibnamefont {Baas}},
  \bibinfo {author} {\bibfnamefont {J.~A.~W.}\ \bibnamefont {van Houwelingen}},
  \bibinfo {author} {\bibfnamefont {N.}~\bibnamefont {Gisin}}, \ and\ \bibinfo
  {author} {\bibfnamefont {H.}~\bibnamefont {Zbinden}},\ }\href@noop {}
  {\bibfield  {journal} {\bibinfo  {journal} {Phys. Rev. Lett.}\ }\textbf
  {\bibinfo {volume} {100}},\ \bibinfo {pages} {220404} (\bibinfo {year}
  {2008})}\BibitemShut {NoStop}%
\bibitem [{\citenamefont {Hensen}\ \emph {et~al.}(2015)\citenamefont {Hensen},
  \citenamefont {Bernien}, \citenamefont {Dr{\'e}au}, \citenamefont {Reiserer},
  \citenamefont {Kalb}, \citenamefont {Blok}, \citenamefont {Ruitenberg},
  \citenamefont {Vermeulen}, \citenamefont {Schouten}, \citenamefont
  {Abell{\'a}n} \emph {et~al.}}]{loophole}%
  \BibitemOpen
  \bibfield  {author} {\bibinfo {author} {\bibfnamefont {B.}~\bibnamefont
  {Hensen}}, \bibinfo {author} {\bibfnamefont {H.}~\bibnamefont {Bernien}},
  \bibinfo {author} {\bibfnamefont {A.}~\bibnamefont {Dr{\'e}au}}, \bibinfo
  {author} {\bibfnamefont {A.}~\bibnamefont {Reiserer}}, \bibinfo {author}
  {\bibfnamefont {N.}~\bibnamefont {Kalb}}, \bibinfo {author} {\bibfnamefont
  {M.}~\bibnamefont {Blok}}, \bibinfo {author} {\bibfnamefont {J.}~\bibnamefont
  {Ruitenberg}}, \bibinfo {author} {\bibfnamefont {R.}~\bibnamefont
  {Vermeulen}}, \bibinfo {author} {\bibfnamefont {R.}~\bibnamefont {Schouten}},
  \bibinfo {author} {\bibfnamefont {C.}~\bibnamefont {Abell{\'a}n}},  \emph
  {et~al.},\ }\href@noop {} {\bibfield  {journal} {\bibinfo  {journal}
  {Nature}\ }\textbf {\bibinfo {volume} {526}},\ \bibinfo {pages} {682}
  (\bibinfo {year} {2015})}\BibitemShut {NoStop}%
\bibitem [{\citenamefont {{Ansmann}}\ \emph {et~al.}(2009)\citenamefont
  {{Ansmann}}, \citenamefont {{Wang}}, \citenamefont {{Bialczak}},
  \citenamefont {{Hofheinz}}, \citenamefont {{Lucero}}, \citenamefont
  {{Neeley}}, \citenamefont {{O'Connell}}, \citenamefont {{Sank}},
  \citenamefont {{Weides}}, \citenamefont {{Wenner}}, \citenamefont
  {{Cleland}},\ and\ \citenamefont {{Martinis}}}]{test11}%
  \BibitemOpen
  \bibfield  {author} {\bibinfo {author} {\bibfnamefont {M.}~\bibnamefont
  {{Ansmann}}}, \bibinfo {author} {\bibfnamefont {H.}~\bibnamefont {{Wang}}},
  \bibinfo {author} {\bibfnamefont {R.~C.}\ \bibnamefont {{Bialczak}}},
  \bibinfo {author} {\bibfnamefont {M.}~\bibnamefont {{Hofheinz}}}, \bibinfo
  {author} {\bibfnamefont {E.}~\bibnamefont {{Lucero}}}, \bibinfo {author}
  {\bibfnamefont {M.}~\bibnamefont {{Neeley}}}, \bibinfo {author}
  {\bibfnamefont {A.~D.}\ \bibnamefont {{O'Connell}}}, \bibinfo {author}
  {\bibfnamefont {D.}~\bibnamefont {{Sank}}}, \bibinfo {author} {\bibfnamefont
  {M.}~\bibnamefont {{Weides}}}, \bibinfo {author} {\bibfnamefont
  {J.}~\bibnamefont {{Wenner}}}, \bibinfo {author} {\bibfnamefont {A.~N.}\
  \bibnamefont {{Cleland}}}, \ and\ \bibinfo {author} {\bibfnamefont {J.~M.}\
  \bibnamefont {{Martinis}}},\ }\href {\doibase 10.1038/nature08363} {\bibfield
   {journal} {\bibinfo  {journal} {Nature}\ }\textbf {\bibinfo {volume}
  {461}},\ \bibinfo {pages} {504} (\bibinfo {year} {2009})}\BibitemShut
  {NoStop}%
\bibitem [{\citenamefont {Lamehi-Rachti}\ and\ \citenamefont
  {Mittig}(1976)}]{PhysRevD.14.2543}%
  \BibitemOpen
  \bibfield  {author} {\bibinfo {author} {\bibfnamefont {M.}~\bibnamefont
  {Lamehi-Rachti}}\ and\ \bibinfo {author} {\bibfnamefont {W.}~\bibnamefont
  {Mittig}},\ }\href {\doibase 10.1103/PhysRevD.14.2543} {\bibfield  {journal}
  {\bibinfo  {journal} {Phys. Rev. D}\ }\textbf {\bibinfo {volume} {14}},\
  \bibinfo {pages} {2543} (\bibinfo {year} {1976})}\BibitemShut {NoStop}%
\bibitem [{\citenamefont {Rosenfeld}\ \emph {et~al.}(2017)\citenamefont
  {Rosenfeld}, \citenamefont {Burchardt}, \citenamefont {Garthoff},
  \citenamefont {Redeker}, \citenamefont {Ortegel}, \citenamefont {Rau},\ and\
  \citenamefont {Weinfurter}}]{PhysRevLett.119.010402}%
  \BibitemOpen
  \bibfield  {author} {\bibinfo {author} {\bibfnamefont {W.}~\bibnamefont
  {Rosenfeld}}, \bibinfo {author} {\bibfnamefont {D.}~\bibnamefont
  {Burchardt}}, \bibinfo {author} {\bibfnamefont {R.}~\bibnamefont {Garthoff}},
  \bibinfo {author} {\bibfnamefont {K.}~\bibnamefont {Redeker}}, \bibinfo
  {author} {\bibfnamefont {N.}~\bibnamefont {Ortegel}}, \bibinfo {author}
  {\bibfnamefont {M.}~\bibnamefont {Rau}}, \ and\ \bibinfo {author}
  {\bibfnamefont {H.}~\bibnamefont {Weinfurter}},\ }\href {\doibase
  10.1103/PhysRevLett.119.010402} {\bibfield  {journal} {\bibinfo  {journal}
  {Phys. Rev. Lett.}\ }\textbf {\bibinfo {volume} {119}},\ \bibinfo {pages}
  {010402} (\bibinfo {year} {2017})}\BibitemShut {NoStop}%
\bibitem [{\citenamefont {Schmied}\ \emph {et~al.}(2016)\citenamefont
  {Schmied}, \citenamefont {Bancal}, \citenamefont {Allard}, \citenamefont
  {Fadel}, \citenamefont {Scarani}, \citenamefont {Treutlein},\ and\
  \citenamefont {Sangouard}}]{schmied2016bell}%
  \BibitemOpen
  \bibfield  {author} {\bibinfo {author} {\bibfnamefont {R.}~\bibnamefont
  {Schmied}}, \bibinfo {author} {\bibfnamefont {J.-D.}\ \bibnamefont {Bancal}},
  \bibinfo {author} {\bibfnamefont {B.}~\bibnamefont {Allard}}, \bibinfo
  {author} {\bibfnamefont {M.}~\bibnamefont {Fadel}}, \bibinfo {author}
  {\bibfnamefont {V.}~\bibnamefont {Scarani}}, \bibinfo {author} {\bibfnamefont
  {P.}~\bibnamefont {Treutlein}}, \ and\ \bibinfo {author} {\bibfnamefont
  {N.}~\bibnamefont {Sangouard}},\ }\href@noop {} {\bibfield  {journal}
  {\bibinfo  {journal} {Science}\ }\textbf {\bibinfo {volume} {352}},\ \bibinfo
  {pages} {441} (\bibinfo {year} {2016})}\BibitemShut {NoStop}%
\bibitem [{\citenamefont {Shin}\ \emph {et~al.}(2019)\citenamefont {Shin},
  \citenamefont {Henson}, \citenamefont {Hodgman}, \citenamefont {Wasak},
  \citenamefont {Chwede\'nczuk},\ and\ \citenamefont
  {Truscott}}]{shin2019bell}%
  \BibitemOpen
  \bibfield  {author} {\bibinfo {author} {\bibfnamefont {D.~K.}\ \bibnamefont
  {Shin}}, \bibinfo {author} {\bibfnamefont {B.~M.}\ \bibnamefont {Henson}},
  \bibinfo {author} {\bibfnamefont {S.~S.}\ \bibnamefont {Hodgman}}, \bibinfo
  {author} {\bibfnamefont {T.}~\bibnamefont {Wasak}}, \bibinfo {author}
  {\bibfnamefont {J.}~\bibnamefont {Chwede\'nczuk}}, \ and\ \bibinfo {author}
  {\bibfnamefont {A.~G.}\ \bibnamefont {Truscott}},\ }\href@noop {} {\bibfield
  {journal} {\bibinfo  {journal} {Nature Communications}\ }\textbf {\bibinfo
  {volume} {10}},\ \bibinfo {pages} {4447} (\bibinfo {year}
  {2019})}\BibitemShut {NoStop}%
\bibitem [{\citenamefont {Gross}\ \emph {et~al.}(2010)\citenamefont {Gross},
  \citenamefont {Zibold}, \citenamefont {Nicklas}, \citenamefont {Esteve},\
  and\ \citenamefont {Oberthaler}}]{gross2010nonlinear}%
  \BibitemOpen
  \bibfield  {author} {\bibinfo {author} {\bibfnamefont {C.}~\bibnamefont
  {Gross}}, \bibinfo {author} {\bibfnamefont {T.}~\bibnamefont {Zibold}},
  \bibinfo {author} {\bibfnamefont {E.}~\bibnamefont {Nicklas}}, \bibinfo
  {author} {\bibfnamefont {J.}~\bibnamefont {Esteve}}, \ and\ \bibinfo {author}
  {\bibfnamefont {M.~K.}\ \bibnamefont {Oberthaler}},\ }\href@noop {}
  {\bibfield  {journal} {\bibinfo  {journal} {Nature}\ }\textbf {\bibinfo
  {volume} {464}},\ \bibinfo {pages} {1165} (\bibinfo {year}
  {2010})}\BibitemShut {NoStop}%
\bibitem [{\citenamefont {Riedel}\ \emph {et~al.}(2010)\citenamefont {Riedel},
  \citenamefont {B{\"o}hi}, \citenamefont {Li}, \citenamefont {H{\"a}nsch},
  \citenamefont {Sinatra},\ and\ \citenamefont {Treutlein}}]{riedel2010atom}%
  \BibitemOpen
  \bibfield  {author} {\bibinfo {author} {\bibfnamefont {M.~F.}\ \bibnamefont
  {Riedel}}, \bibinfo {author} {\bibfnamefont {P.}~\bibnamefont {B{\"o}hi}},
  \bibinfo {author} {\bibfnamefont {Y.}~\bibnamefont {Li}}, \bibinfo {author}
  {\bibfnamefont {T.~W.}\ \bibnamefont {H{\"a}nsch}}, \bibinfo {author}
  {\bibfnamefont {A.}~\bibnamefont {Sinatra}}, \ and\ \bibinfo {author}
  {\bibfnamefont {P.}~\bibnamefont {Treutlein}},\ }\href@noop {} {\bibfield
  {journal} {\bibinfo  {journal} {Nature}\ }\textbf {\bibinfo {volume} {464}},\
  \bibinfo {pages} {1170} (\bibinfo {year} {2010})}\BibitemShut {NoStop}%
\bibitem [{\citenamefont {Leroux}\ \emph {et~al.}(2010)\citenamefont {Leroux},
  \citenamefont {Schleier-Smith},\ and\ \citenamefont
  {Vuleti{\'c}}}]{leroux2010orientation}%
  \BibitemOpen
  \bibfield  {author} {\bibinfo {author} {\bibfnamefont {I.~D.}\ \bibnamefont
  {Leroux}}, \bibinfo {author} {\bibfnamefont {M.~H.}\ \bibnamefont
  {Schleier-Smith}}, \ and\ \bibinfo {author} {\bibfnamefont {V.}~\bibnamefont
  {Vuleti{\'c}}},\ }\href@noop {} {\bibfield  {journal} {\bibinfo  {journal}
  {Phys. Rev. Lett.}\ }\textbf {\bibinfo {volume} {104}},\ \bibinfo {pages}
  {250801} (\bibinfo {year} {2010})}\BibitemShut {NoStop}%
\bibitem [{\citenamefont {Chen}\ \emph {et~al.}(2011)\citenamefont {Chen},
  \citenamefont {Bohnet}, \citenamefont {Sankar}, \citenamefont {Dai},\ and\
  \citenamefont {Thompson}}]{chen2011conditional}%
  \BibitemOpen
  \bibfield  {author} {\bibinfo {author} {\bibfnamefont {Z.}~\bibnamefont
  {Chen}}, \bibinfo {author} {\bibfnamefont {J.~G.}\ \bibnamefont {Bohnet}},
  \bibinfo {author} {\bibfnamefont {S.~R.}\ \bibnamefont {Sankar}}, \bibinfo
  {author} {\bibfnamefont {J.}~\bibnamefont {Dai}}, \ and\ \bibinfo {author}
  {\bibfnamefont {J.~K.}\ \bibnamefont {Thompson}},\ }\href@noop {} {\bibfield
  {journal} {\bibinfo  {journal} {Phys. Rev. Lett.}\ }\textbf {\bibinfo
  {volume} {106}},\ \bibinfo {pages} {133601} (\bibinfo {year}
  {2011})}\BibitemShut {NoStop}%
\bibitem [{\citenamefont {Esteve}\ \emph {et~al.}(2008)\citenamefont {Esteve},
  \citenamefont {Gross}, \citenamefont {Weller}, \citenamefont {Giovanazzi},\
  and\ \citenamefont {Oberthaler}}]{esteve2008squeezing}%
  \BibitemOpen
  \bibfield  {author} {\bibinfo {author} {\bibfnamefont {J.}~\bibnamefont
  {Esteve}}, \bibinfo {author} {\bibfnamefont {C.}~\bibnamefont {Gross}},
  \bibinfo {author} {\bibfnamefont {A.}~\bibnamefont {Weller}}, \bibinfo
  {author} {\bibfnamefont {S.}~\bibnamefont {Giovanazzi}}, \ and\ \bibinfo
  {author} {\bibfnamefont {M.}~\bibnamefont {Oberthaler}},\ }\href@noop {}
  {\bibfield  {journal} {\bibinfo  {journal} {Nature}\ }\textbf {\bibinfo
  {volume} {455}},\ \bibinfo {pages} {1216} (\bibinfo {year}
  {2008})}\BibitemShut {NoStop}%
\bibitem [{\citenamefont {Strobel}\ \emph {et~al.}(2014)\citenamefont
  {Strobel}, \citenamefont {Muessel}, \citenamefont {Linnemann}, \citenamefont
  {Zibold}, \citenamefont {Hume}, \citenamefont {Pezz\'e}, \citenamefont
  {Smerzi},\ and\ \citenamefont {Oberthaler}}]{smerzi_ob}%
  \BibitemOpen
  \bibfield  {author} {\bibinfo {author} {\bibfnamefont {H.}~\bibnamefont
  {Strobel}}, \bibinfo {author} {\bibfnamefont {W.}~\bibnamefont {Muessel}},
  \bibinfo {author} {\bibfnamefont {D.}~\bibnamefont {Linnemann}}, \bibinfo
  {author} {\bibfnamefont {T.}~\bibnamefont {Zibold}}, \bibinfo {author}
  {\bibfnamefont {D.~B.}\ \bibnamefont {Hume}}, \bibinfo {author}
  {\bibfnamefont {L.}~\bibnamefont {Pezz\'e}}, \bibinfo {author} {\bibfnamefont
  {A.}~\bibnamefont {Smerzi}}, \ and\ \bibinfo {author} {\bibfnamefont {M.~K.}\
  \bibnamefont {Oberthaler}},\ }\href {\doibase 10.1126/science.1250147}
  {\bibfield  {journal} {\bibinfo  {journal} {Science}\ }\textbf {\bibinfo
  {volume} {345}},\ \bibinfo {pages} {424} (\bibinfo {year}
  {2014})}\BibitemShut {NoStop}%
\bibitem [{\citenamefont {Braunstein}\ and\ \citenamefont
  {Caves}(1994)}]{braunstein1994statistical}%
  \BibitemOpen
  \bibfield  {author} {\bibinfo {author} {\bibfnamefont {S.~L.}\ \bibnamefont
  {Braunstein}}\ and\ \bibinfo {author} {\bibfnamefont {C.~M.}\ \bibnamefont
  {Caves}},\ }\href@noop {} {\bibfield  {journal} {\bibinfo  {journal} {Phys.
  Rev. Lett.}\ }\textbf {\bibinfo {volume} {72}},\ \bibinfo {pages} {3439}
  (\bibinfo {year} {1994})}\BibitemShut {NoStop}%
\bibitem [{\citenamefont {Brush}(1967)}]{RevModPhys.39.883}%
  \BibitemOpen
  \bibfield  {author} {\bibinfo {author} {\bibfnamefont {S.~G.}\ \bibnamefont
  {Brush}},\ }\href@noop {} {\bibfield  {journal} {\bibinfo  {journal} {Rev.
  Mod. Phys.}\ }\textbf {\bibinfo {volume} {39}},\ \bibinfo {pages} {883}
  (\bibinfo {year} {1967})}\BibitemShut {NoStop}%
\bibitem [{\citenamefont {Baxter}(2016)}]{baxter2016exactly}%
  \BibitemOpen
  \bibfield  {author} {\bibinfo {author} {\bibfnamefont {R.~J.}\ \bibnamefont
  {Baxter}},\ }\href@noop {} {\emph {\bibinfo {title} {Exactly solved models in
  statistical mechanics}}}\ (\bibinfo  {publisher} {Elsevier},\ \bibinfo {year}
  {2016})\BibitemShut {NoStop}%
\bibitem [{\citenamefont {Dalfovo}\ \emph {et~al.}(1999)\citenamefont
  {Dalfovo}, \citenamefont {Giorgini}, \citenamefont {Pitaevskii},\ and\
  \citenamefont {Stringari}}]{dalfovo1999theory}%
  \BibitemOpen
  \bibfield  {author} {\bibinfo {author} {\bibfnamefont {F.}~\bibnamefont
  {Dalfovo}}, \bibinfo {author} {\bibfnamefont {S.}~\bibnamefont {Giorgini}},
  \bibinfo {author} {\bibfnamefont {L.~P.}\ \bibnamefont {Pitaevskii}}, \ and\
  \bibinfo {author} {\bibfnamefont {S.}~\bibnamefont {Stringari}},\ }\href@noop
  {} {\bibfield  {journal} {\bibinfo  {journal} {Reviews of Modern Physics}\
  }\textbf {\bibinfo {volume} {71}},\ \bibinfo {pages} {463} (\bibinfo {year}
  {1999})}\BibitemShut {NoStop}%
\bibitem [{\citenamefont {Shin}\ \emph {et~al.}(2004)\citenamefont {Shin},
  \citenamefont {Saba}, \citenamefont {Pasquini}, \citenamefont {Ketterle},
  \citenamefont {Pritchard},\ and\ \citenamefont {Leanhardt}}]{shin2004atom}%
  \BibitemOpen
  \bibfield  {author} {\bibinfo {author} {\bibfnamefont {Y.}~\bibnamefont
  {Shin}}, \bibinfo {author} {\bibfnamefont {M.}~\bibnamefont {Saba}}, \bibinfo
  {author} {\bibfnamefont {T.}~\bibnamefont {Pasquini}}, \bibinfo {author}
  {\bibfnamefont {W.}~\bibnamefont {Ketterle}}, \bibinfo {author}
  {\bibfnamefont {D.}~\bibnamefont {Pritchard}}, \ and\ \bibinfo {author}
  {\bibfnamefont {A.}~\bibnamefont {Leanhardt}},\ }\href@noop {} {\bibfield
  {journal} {\bibinfo  {journal} {Physical review letters}\ }\textbf {\bibinfo
  {volume} {92}},\ \bibinfo {pages} {050405} (\bibinfo {year}
  {2004})}\BibitemShut {NoStop}%
\bibitem [{\citenamefont {Gati}\ \emph {et~al.}(2006)\citenamefont {Gati},
  \citenamefont {Hemmerling}, \citenamefont {F{\"o}lling}, \citenamefont
  {Albiez},\ and\ \citenamefont {Oberthaler}}]{gati2006noise}%
  \BibitemOpen
  \bibfield  {author} {\bibinfo {author} {\bibfnamefont {R.}~\bibnamefont
  {Gati}}, \bibinfo {author} {\bibfnamefont {B.}~\bibnamefont {Hemmerling}},
  \bibinfo {author} {\bibfnamefont {J.}~\bibnamefont {F{\"o}lling}}, \bibinfo
  {author} {\bibfnamefont {M.}~\bibnamefont {Albiez}}, \ and\ \bibinfo {author}
  {\bibfnamefont {M.~K.}\ \bibnamefont {Oberthaler}},\ }\href@noop {}
  {\bibfield  {journal} {\bibinfo  {journal} {Phys. Rev. Lett.}\ }\textbf
  {\bibinfo {volume} {96}},\ \bibinfo {pages} {130404} (\bibinfo {year}
  {2006})}\BibitemShut {NoStop}%
\bibitem [{\citenamefont {Cavalcanti}\ \emph {et~al.}(2007)\citenamefont
  {Cavalcanti}, \citenamefont {Foster}, \citenamefont {Reid},\ and\
  \citenamefont {Drummond}}]{cavalcanti2007bell}%
  \BibitemOpen
  \bibfield  {author} {\bibinfo {author} {\bibfnamefont {E.~G.}\ \bibnamefont
  {Cavalcanti}}, \bibinfo {author} {\bibfnamefont {C.~J.}\ \bibnamefont
  {Foster}}, \bibinfo {author} {\bibfnamefont {M.~D.}\ \bibnamefont {Reid}}, \
  and\ \bibinfo {author} {\bibfnamefont {P.~D.}\ \bibnamefont {Drummond}},\
  }\href@noop {} {\bibfield  {journal} {\bibinfo  {journal} {Phys. Rev. Lett.}\
  }\textbf {\bibinfo {volume} {99}},\ \bibinfo {pages} {210405} (\bibinfo
  {year} {2007})}\BibitemShut {NoStop}%
\bibitem [{\citenamefont {He}\ \emph {et~al.}(2011)\citenamefont {He},
  \citenamefont {Drummond},\ and\ \citenamefont {Reid}}]{he2011entanglement}%
  \BibitemOpen
  \bibfield  {author} {\bibinfo {author} {\bibfnamefont {Q.}~\bibnamefont
  {He}}, \bibinfo {author} {\bibfnamefont {P.}~\bibnamefont {Drummond}}, \ and\
  \bibinfo {author} {\bibfnamefont {M.}~\bibnamefont {Reid}},\ }\href@noop {}
  {\bibfield  {journal} {\bibinfo  {journal} {Physical Review A}\ }\textbf
  {\bibinfo {volume} {83}},\ \bibinfo {pages} {032120} (\bibinfo {year}
  {2011})}\BibitemShut {NoStop}%
\bibitem [{\citenamefont {Cavalcanti}\ \emph {et~al.}(2011)\citenamefont
  {Cavalcanti}, \citenamefont {He}, \citenamefont {Reid},\ and\ \citenamefont
  {Wiseman}}]{cavalcanti2011unified}%
  \BibitemOpen
  \bibfield  {author} {\bibinfo {author} {\bibfnamefont {E.}~\bibnamefont
  {Cavalcanti}}, \bibinfo {author} {\bibfnamefont {Q.}~\bibnamefont {He}},
  \bibinfo {author} {\bibfnamefont {M.}~\bibnamefont {Reid}}, \ and\ \bibinfo
  {author} {\bibfnamefont {H.}~\bibnamefont {Wiseman}},\ }\href@noop {}
  {\bibfield  {journal} {\bibinfo  {journal} {Physical Review A}\ }\textbf
  {\bibinfo {volume} {84}},\ \bibinfo {pages} {032115} (\bibinfo {year}
  {2011})}\BibitemShut {NoStop}%
\bibitem [{\citenamefont {Niezgoda}\ \emph {et~al.}(2020)\citenamefont
  {Niezgoda}, \citenamefont {Panfil},\ and\ \citenamefont
  {Chwede\ifmmode~\acute{n}\else \'{n}\fi{}czuk}}]{spiny.milosz}%
  \BibitemOpen
  \bibfield  {author} {\bibinfo {author} {\bibfnamefont {A.}~\bibnamefont
  {Niezgoda}}, \bibinfo {author} {\bibfnamefont {M.}~\bibnamefont {Panfil}}, \
  and\ \bibinfo {author} {\bibfnamefont {J.}~\bibnamefont
  {Chwede\ifmmode~\acute{n}\else \'{n}\fi{}czuk}},\ }\href {\doibase
  10.1103/PhysRevA.102.042206} {\bibfield  {journal} {\bibinfo  {journal}
  {Phys. Rev. A}\ }\textbf {\bibinfo {volume} {102}},\ \bibinfo {pages}
  {042206} (\bibinfo {year} {2020})}\BibitemShut {NoStop}%
\bibitem [{\citenamefont {Dziarmaga}\ \emph {et~al.}(2002)\citenamefont
  {Dziarmaga}, \citenamefont {Smerzi}, \citenamefont {Zurek},\ and\
  \citenamefont {Bishop}}]{dziarmaga2002dynamics}%
  \BibitemOpen
  \bibfield  {author} {\bibinfo {author} {\bibfnamefont {J.}~\bibnamefont
  {Dziarmaga}}, \bibinfo {author} {\bibfnamefont {A.}~\bibnamefont {Smerzi}},
  \bibinfo {author} {\bibfnamefont {W.}~\bibnamefont {Zurek}}, \ and\ \bibinfo
  {author} {\bibfnamefont {A.}~\bibnamefont {Bishop}},\ }\href@noop {}
  {\bibfield  {journal} {\bibinfo  {journal} {Physical review letters}\
  }\textbf {\bibinfo {volume} {88}},\ \bibinfo {pages} {167001} (\bibinfo
  {year} {2002})}\BibitemShut {NoStop}%
\bibitem [{\citenamefont {Trenkwalder}\ \emph {et~al.}(2016)\citenamefont
  {Trenkwalder}, \citenamefont {Spagnolli}, \citenamefont {Semeghini},
  \citenamefont {Coop}, \citenamefont {Landini}, \citenamefont {Castilho},
  \citenamefont {Pezze}, \citenamefont {Modugno}, \citenamefont {Inguscio},
  \citenamefont {Smerzi} \emph {et~al.}}]{trenkwalder2016quantum}%
  \BibitemOpen
  \bibfield  {author} {\bibinfo {author} {\bibfnamefont {A.}~\bibnamefont
  {Trenkwalder}}, \bibinfo {author} {\bibfnamefont {G.}~\bibnamefont
  {Spagnolli}}, \bibinfo {author} {\bibfnamefont {G.}~\bibnamefont
  {Semeghini}}, \bibinfo {author} {\bibfnamefont {S.}~\bibnamefont {Coop}},
  \bibinfo {author} {\bibfnamefont {M.}~\bibnamefont {Landini}}, \bibinfo
  {author} {\bibfnamefont {P.}~\bibnamefont {Castilho}}, \bibinfo {author}
  {\bibfnamefont {L.}~\bibnamefont {Pezze}}, \bibinfo {author} {\bibfnamefont
  {G.}~\bibnamefont {Modugno}}, \bibinfo {author} {\bibfnamefont
  {M.}~\bibnamefont {Inguscio}}, \bibinfo {author} {\bibfnamefont
  {A.}~\bibnamefont {Smerzi}},  \emph {et~al.},\ }\href@noop {} {\bibfield
  {journal} {\bibinfo  {journal} {Nature physics}\ }\textbf {\bibinfo {volume}
  {12}},\ \bibinfo {pages} {826} (\bibinfo {year} {2016})}\BibitemShut
  {NoStop}%
\end{thebibliography}
\end{document}